# Antibacterial and antifungal activity of various extracts of Bacopa monnieri

**FAZLUL MKK[1], DEEPTHI SP[2], MOHAMMED IRFAN[3], FARZANA Y[4], MUNIRA B[5], NAZMUL MHM[6*]**

[1]Faculty of Industrial Sciences & Technology, Universiti Malaysia Pahang, Gambang, 26300 Pahang, Malaysia

[2]Faculty of Medicine, MAHSA University, Bandar Saujana Putra campus, 42610 Kuala Langat, Selangor, Malaysia

[3]Faculty of Medicine, International Medical University, Bukit Jalil 57000, Kuala Lumpur, Malaysia

[4]Faculty of Science, Lincoln University, 12-18, Jalan SS6/12, Off Jalan Perbandaran, 47301 Petaling Jaya, Selangor Malaysia

[5]Faculty of Medicine, SEGi University, Kota Damansara, Selangor, Malaysia

[6]Graduate School of Medicine, Perdana University, Jalan MAEPS Perdana, Serdang, 43400 Selangor, Malaysia E-mail: poorpiku@yahoo.com

Received: 09.11.18, Revised: 09.12.18, Accepted: 09.01.19

## ABSTRACT

Bacopa monnieri was also known as “Brahmi.” Theses plant has been considered as medicinal plants in Ayurvedic systems for centuries. This present study was carried out to evaluate the antibacterial and antifungal properties of ethanolic, diethyl ether, ethyl acetate, and aqueous extracts of B. monnieri. Agar disc diffusion tests were carried out to determine the antimicrobial effects of ethanolic, diethyl ether, ethyl acetate and aqueous extracts of B. monnieri against Gram-positive (Staphylococcus aureus ATCC 25923), Gram-negative (Escherichia coli ATCC 25922) bacterial strains and antifungal strains (Aspergillus flavus, and Candida albicans). Among the various extracts, diethyl ether extracts of B. monnieri has an antibacterial potency against Staphylococcus aureus (gram positive), and ethyl acetate extract showed effects on E. coli (gram negative) at higher concentrations of 300 µg mL-1. The ethanolic extract has potent antifungal activity against the fungus (Aspergillus flavus, and Candida albicans) compared to diethyl ether and ethyl acetate-ether. Both extracts (diethyl ether and ethyl acetate) has a minimum antifungal effect while these extracts showed more inhibitory effects on tested bacteria. Inhibitory effects of the aqueous extract were not observed at all concentration (100 µg mL-1, 200 µg mL-1and 300 µg mL-1) against the entire examined bacterial and fungal species. The finding of this study indicates the antibacterial and antifungal activity exhibited by the ethanolic, diethyl ether, and ethyl acetate extract of B. monnieri can be validated to the management of medicinal plant throughout the world.

## INTRODUCTION

Medicinal herbs have been considered safe, effective, and play an essential role in modern pharmacology (1). B. monnieri is a creeping, succulent, herb known as an essential medicinal plant throughout the world (2). B. monnieri is used medicinally for the treatment of insanity, epilepsy, and skin disease, etc. (3). This plant has reported as a sedative, vasoconstrictor, antimicrobial, antifungal, anti-inflammatory, antiepileptic, and anthelmintic activities (3). B. monnieri, also known as Brahmi in the Ayurvedic medicine for centuries (4). Antimicrobial properties of several parts of the plant were included in this study. The entire plant is being used as a nervous tonic (5). Identification, separation, quantification, and standardization of B. monnieri compounds was carried out over the years. Explore antimicrobial compound in the plants is an important study (6). Pharmacological effects of B. monnieri is due to its active compounds of alkaloids, sterols, saponins, hersaponin, stigmasterol, beta-sitosterol, and bacopasaponins etc. (7).Bacterial and fungal infection are severe for immunologically deficient patients; an urgent need for natural products may be useful and control multidrug resistance in microbial activities. Therefore, the reason of the present study was to assess and investigate the B. monnieri extracts effects against the different tested microbial (bacteria and fungus) strains. This activity supports in the treatment of infections caused by various microorganisms. Infections due to microbes pose a health problem around the world and their resistance mechanisms have contributed to worse clinical outcomes. It is noteworthy to mention that plants extracts are a possible source of antimicrobial agents, alternative to cheap, and effective herbal drugs to treat many disorders and diseases against common bacterial infections.

## Materials and Methods

### Plant material

B. monnieri (whole plant) were selected from wild vegetations in Selangor, Malaysia for the study. The plant material was confirmed and authenticated at Perdana University, Malaysia.

### Preparation of various extract

The 250 grams fresh aerial parts of the plant's materials were cleaned by removing dead stems/roots/leaves or any other unwanted materials and air-dried for 20 days. A measured quantity of 20 grams of dried powder was soaked in 200 ml ethanol, diethyl ether, ethyl acetate and aqueous in a round bottom flask at room temperature for 24 hours. Each extract was passed through filter paper no1 (Whatman, England) at the end of extraction. At room temperature, the filtrate was kept dry until extract dried completely (8). The weight of the dried extract was 0.78 grams. The yield of the extract obtained was 3.20% (w/w). All the extracts were stored in containers (airtight) at 48ºC for additional testing.

### Determination of Antibacterial and Antifungal activities

### Culture Media

The antibacterial test was carried out by Nutrient Agar (NA) and Muller Hinton broth (MHB). Sabaroud's Dextrose Agar (SDA) was used for fungal culture and susceptibility test.

### Inoculum

The bacteria were inoculated into Muller Hinton broth (MHB and incubated at 370C for 4 hours and confirm suspension for approximately 105 CFU/ml. Similarly, fungal strains were inoculated into Sabouraud's dextrose broth for 6 hours.

### Microorganisms

The antibacterial activity of ethanolic, diethyl ether, ethyl acetate, and aqueous extract was determined by individually testing on Gram-positive and Gram-negative bacterial strains. The Gram-positive (Staphylococcus aureus ATCC 25923) and Gram-negative (Escherichia coli ATCC 25922) strains were used. All the strains were cultured and maintained on nutrient agar at 4°C. Two types of medically important fungi (Aspergillus flavus, and Candida albicans) were selected which causes infection in human. Potential antibacterial and antifungal properties of B. monnieri extracts were measured in terms of zone of inhibition.

### Drug used

Antibacterial (Ciprofloxacin 5 gm/ml) and antifungal (clotrimazole 5 gm/ml) were used as positive standard for tested microorganisms.

### Antibacterial activity assay

The antibacterial activity was determined by agar disc diffusion (9-10). A volume of 250 µl of the bacterial culture (105CFU) was plated on the Nutrient Agar (NA) plate and left it to dry. Ethanolic, diethyl ether, ethyl acetate, and aqueous acetate extract on 6 mm filter paper discs were loaded for the concentration (0.1mL, 0.2mL, and 0.3mL) and kept for completely dry. Ciprofloxacin discs (5mg/ml) was used as reference standard control. The plates were incubated overnight at 37°C. The zone of inhibition (mm) on the surface of each test organism around the paper disc were determined according to the Clinical and Laboratory Standards Institute for their antibacterial activity (11). The average diameters of zones were calculated after each test for 3 times.

### Antifungal sensitivity test

All the extracts of B. monnieri were then tested for their fungal toxicity against Aspergillus flavus and Candida albicans. These extracts were added in SDA agar medium for the concentration of 0.1mL, 0.2mL, and 0.3mL. Extracts added SDA medium was solidified in Petri plates. Hardened 10 mm SDA agar block was cut and placed at the center of the actively growing Aspergillus flavus and Candia albicans petri plates for incubation at 30°C for 7-14 days (8). The average diameters were measured. Clotrimazol used as positive control standards.

## Statistical Analysis

All the experiments were repeated for 3 times and obtained data were expressed as standard deviation.

## Results

Ethanol, diethyl ether and ethyl acetate extract of Bacopa monnieri at various concentration showed the different inhibitory effect against bacteria (gram positive and gram negative) as well as fungus. Among various (ethanol, diethyl ether, and ethyl acetate) extract of B. monnieri, diethyl ether showed the maximum (18.36±38mm) inhibitory effect against Staphylococcus aureus at higher concentration (300 $\mu$g). The potency of ethanol 15.16±22 mm and ethyl acetate 8.20±10 mm, respectively against gram-positive bacteria (Staphylococcus aureus). In gram-negative bacteria

(E. coli), various extract (ethanol, diethyl ether, and ethyl acetate) showed the maximum 9.42±17, 11.06±38, and 13.3±28 respectively at higher concentration (300 $\mu$g).The antifungal capability of various extracts of B. monnieri showed a moderate effect over all types of fungus (Aspergillus flavus and Candida albicans). Ethanolic extracts were found to have maximum activity (10.83±25) followed by diethyl ether extract (8.61±14) and ethyl acetate extract (6.66±22). Ethyl acetate extract exhibited lesser inhibitory potency when compared with ethanolic and diethyl ether extract. At the concentration of 300 $\mu$g, all the extracts were more effective against bacteria and fungus compared to less concentration (100 $\mu$g and 200 $\mu$g) (Table 1). The inhibitory effect was not shown by the aqueous extract against both bacteria and fungus studied.

**Table 1: Inhibitory effect of various extracts of Bacopa monnieri (Linn.) against pathogenic organisms**

| Microorganisms | Ethanol ($\mu$g mL$^{-1}$) | | | Diethyl ether ($\mu$g mL$^{-1}$) | | | Ethyl acetate ($\mu$g mL$^{-1}$) | | | Aqueous ($\mu$g mL$^{-1}$) | | |
|---|---|---|---|---|---|---|---|---|---|---|---|---|
| | 100 | 200 | 300 | 100 | 200 | 300 | 100 | 200 | 300 | 100 | 200 | 300 |
| *Staphylococcus aureus* | 8.93±.04 | 10.23 ±14 | 15.16 ±22 | 12.16 ±22 | 14.33 ±67 | 18.36 ±38 | 6.92 ±32 | 7.40±58 | 8.20 ±10 | NA | NA | NA |
| *Escherich coli* | 5.42±28 | 8.60 ±24 | 9.42 ±17 | 9.61 ±14 | 9.80 ±07 | 11.06 ±38 | 11.8 ±67 | 12.06 ±38 | 13.3 ±28 | NA | NA | NA |
| *Aspergillus flavus* | 10.03 ±52 | 10.43 ± 17 | 10.83 ±25 | 7.41 ±88 | 7.55±26 | 8.61±14 | 4.52 ±26 | 5.78±28 | 6.66 ±22 | NA | NA | NA |
| *Candida albicans* | 9.60±22 | 10.5 ±14 | 11.33 ±33 | 6.36 ±36 | 7.36 ±14 | 8.66 ±31 | 4.56 ±43 | 5.23±57 | 6.63 ±38 | NA | NA | NA |

Values are mean of three readings; NA: No activity found at this concentration; Standards: Ciprofloxacin 5mg/ml; Clotrimazole 5mg/ml.

Antimicrobial activities of the various extracts of Bacopa monnieri (Linn.) were tested against 2 pathogenic bacteria and 2 fungi with standard antimicrobial agents (Ciprofloxacin5mg/ml) and antifungal (Clotrimazole 5mg/ml). At the concentration of 300 $\mu$g, various plant extracts of B. monnieri and antimicrobial agents were tested against microorganisms for their susceptibility (Figure 1).

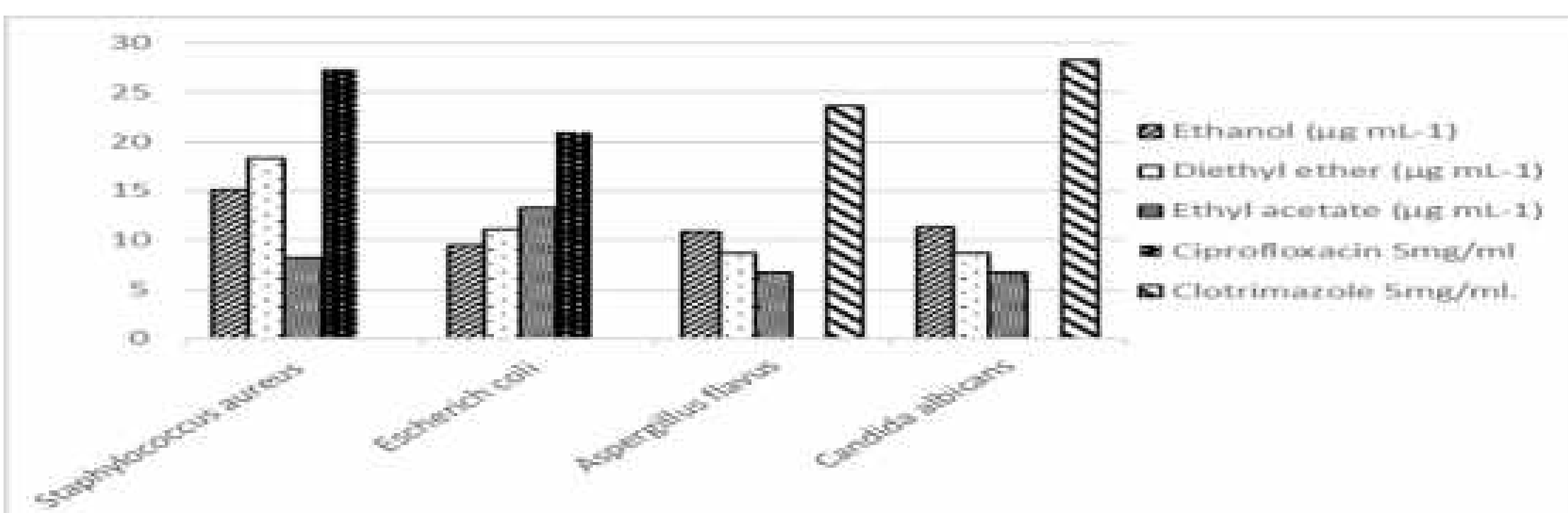

**Fig.1: Susceptibility patterns of both various extract of B. monnieri and antimicrobial agents.**

## Discussion

Antimicrobial activities of various plant sources are existing treatment around the world (12). Numbers of studies have been reported on the antimicrobial properties of herbs, extraction of essential oil, extracts, and decoctions. *Bacopa monnieri* plant properties are anti-infective agents against various types of bacteria and fungus (13). Medicinally, this is also used for inflamed tissues (14). Compounds of *B. monnieri* such as flavonoids, tannins, and phenolic acid has effects on antimicrobial treatment (15). Neurological disorders can be treated with *Bacopa monnieri* (16).Results of the present study showed that the various extract of *B. monnieri* possesses antibacterial properties against human pathogenic organisms, which are in agreement with several studies reported (17-18). Among the different concentration of extracts, diethyl ether extract inhibits *Staphylococcus aureus* growth at higher concentration (300µg). All the extracts of *B. monnieri* possessed antibacterial properties except aqueous extracts showed no inhibitory effects on the tested bacterial species. Gram-positive bacteria were highly affected in comparison to gram-negative bacteria by all the extracts except ethyl acetate. Gram-positive bacteria exhibit a higher susceptibility compared to gram-negative bacteria by various plant extracts. Diethyl ether and ethyl acetate extracts showed higher antibacterial activity against gram-positive (*Staphylococcus aureus*) and gram-negative (*E.coli*) bacteria respectively, which is in agreement with several previous findings (8, 19).On the other hand, ethanolic extracts of *B. monnieri* showed an elevated effect against both the fungal species which is similar to a study by Sampathkumar et al (20). The aqueous extract remained non-responsive or inhibited the growth of all tested fungus. These could be due to the loss of some active compounds during the extraction processes or topographical/environmental effect leads to variation in contents. Besides, the antimicrobial activities of the various extracts of *B. monnieri* were observed and compared with the standard drugs (Ciprofloxacin and Clotrimazole) at a concentration of 300 µg against all the microorganisms investigated.Development of resistance mechanisms is a never-ending process by the microbes against known antibiotics (21-22). Thus new antimicrobial compounds require a huge concern in the medical field. This present study provides evidence that *B. monnieri* is a promising plant which has antimicrobial properties. However, active ingredients and mechanism of action should be enlightened and requires a further test against various types of bacteria and fungus.

## Conclusion

The present study has revealed that the *B. monnieri* plant extracts is a potential antimicrobial agent having a broad-spectrum activity. Thus, plants derived chemicals should be a priority towards sustainable management and appropriate utilization of biodiversity. Further studies on the isolation of pure compounds could exhibit an exciting result.

## Conflict of interest

None

## Acknowledgment

Authors like to appreciate co-others for their unconditional support.